\documentclass[twocolumn]{aastex63}

\hypersetup{linkcolor=red, citecolor=blue}
\usepackage{comment}
\usepackage{amssymb}
\usepackage{amsfonts}
\usepackage{amsmath}
\usepackage[varg]{txfonts}
\usepackage{natbib}
\usepackage{color}
\usepackage{graphicx}
\usepackage{hyperref}

\newcommand{\RNum}[1]{\uppercase\expandafter{\romannumeral #1\relax}}
\newcommand{\felix}[1]{\textsc{F\small{ELIX}}}

\newcommand\aastex{AAS\TeX}

\received{---}
\revised{---}
\accepted{---}
\submitjournal{RNAAS}

\shorttitle{\aastex\ Mode behaviours in the sdB and white dwarf stars}
\shortauthors{Zong et al. 2021}
\begin{document}

\title{Amplitude and frequency modulation of super-Nyquist frequency from {\sl Kepler} photometric sampling}

\email{weikai.zong@bnu.edu.cn}

\author{Weikai Zong}
\affil{Department of Astronomy, Beijing Normal University, Beijing~100875, P.~R.~China}

\author{St\'ephane Charpinet}
\affil{IRAP, Universit\'e de Toulouse, CNRS, UPS, CNES, 14 avenue Edouard Belin, F-31400, Toulouse, France}

\begin{abstract}

We present a simulation showing that super-Nyquist frequencies may have periodic amplitude and frequency modulations, even if actually stable, in time series sampled like the {\sl Kepler} data. These modulations are caused by the barycentric time correction, which destroys the evenly spaced time measurements, making the Nyquist frequency variable over the spacecraft orbit around the Sun. These modulations can easily be identified in pulsating stars from {\sl Kepler}'s photometric data.

\end{abstract}

\keywords{technique: photometric --- stars: frequency --- stars: modulations}

\section{Introduction}
    Pulsating stars, exhibiting periodic luminosity variations, offer the unique opportunity to probe their interiors with the technique of asteroseismology. Their intrinsic frequencies can be measured with high precision from high-quality fast photometry over long time baselines, usually achieved from spaceborne observations. The {\sl Kepler} spacecraft monitored stars in the 105$^\circ$ region near the Cygnus and Lyrae constellations during its prime mission, leading to significant breakthroughs for many types of pulsating stars across the Hertzsprung-Russell diagram \citep{2021RvMP...93a5001A}. Notably, the 4-yr time baseline achieved for various pulsators allowed us to resolve the variability of oscillation modes, providing observations to further develop nonlinear theory of stellar oscillations \citep[see, e.g.,][]{zong2016a}. 

Nevertheless, the delivery of {\sl Kepler} data was limited by downlink bandwidth, which could only provide high cadence ($\Delta t \sim 58.85$\,s) light curves for about 500 stars and long cadence ($\Delta t\sim29.4$\,min) light curves  for about 200,000 stars. Using Fourier transforms, these photometric data are usually transposed into frequency space up to the Nyquist frequency $f_{Ny} = 1/(2 \Delta t)$, where the pulsation signals can easily be identified. However, mode frequencies may sometimes be over the Nyquist limit when fast pulsators are involved. These are called super-Nyquist frequencies. For instance, p-modes in hot B subdwarfs can reach frequencies up to ten thousands $\mu$Hz, which is beyond the Nyquist limit of {\sl Kepler}'s high cadence. Super-Nyquist frequencies can still be detected as a result of a reflection relative to the Nyquist frequency. However, the profile of those super-Nyquist peaks turns out to be complex \citep[see, e.g.,][]{2012MNRAS.424.2686B,2013MNRAS.430.2986M}, thus contaminating measurements of intrinsic amplitude and frequency modulation patterns. In this context, we report a simple simulation of stable signals beyond the Nyquist frequency that clearly show periodic amplitude and frequency modulations as a consequence of the frequency reflection relative to $f_{Ny}$.

\section{Results and discussion} \label{s:r}
The simulations are performed with the following steps: (1) To reproduce {\sl Kepler}'s sampling, we take the time array from one {\sl Kepler} target with long cadence data from Quarter 1 to 17. (2) The flux array is constructed with a process similar to the one described in \citet[][see details in Section 2.3]{zong2016b}, but with injection of three frequencies $f_0 = 14.6880$~c/d, $f_1 = f_0/0.77$ and $f_2 = f_0 + f_1$, where $f_2 > f_{Ny} \sim 24.5$~c/d. To make the measurements significant, their intrinsic amplitudes are stable and of the order of $0.1$\,mag. (3) Amplitude and frequency modulations are measured with the same method described in \citet[][see details in Section 3]{zong2018}, but using a time step and a window of 45 days and 135 days, respectively. 

\begin{figure*}
\centering
 %\epsscale{.80}
\includegraphics[width=8cm]{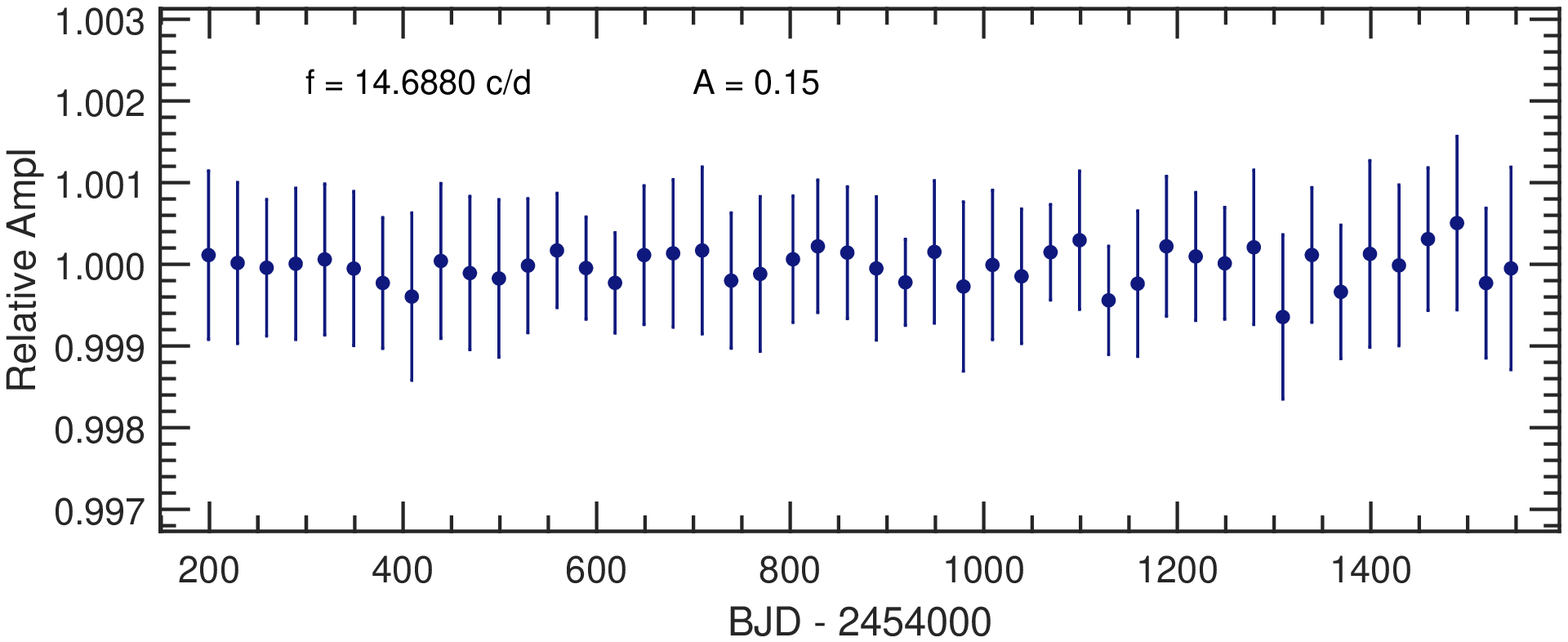}
\includegraphics[width=8cm]{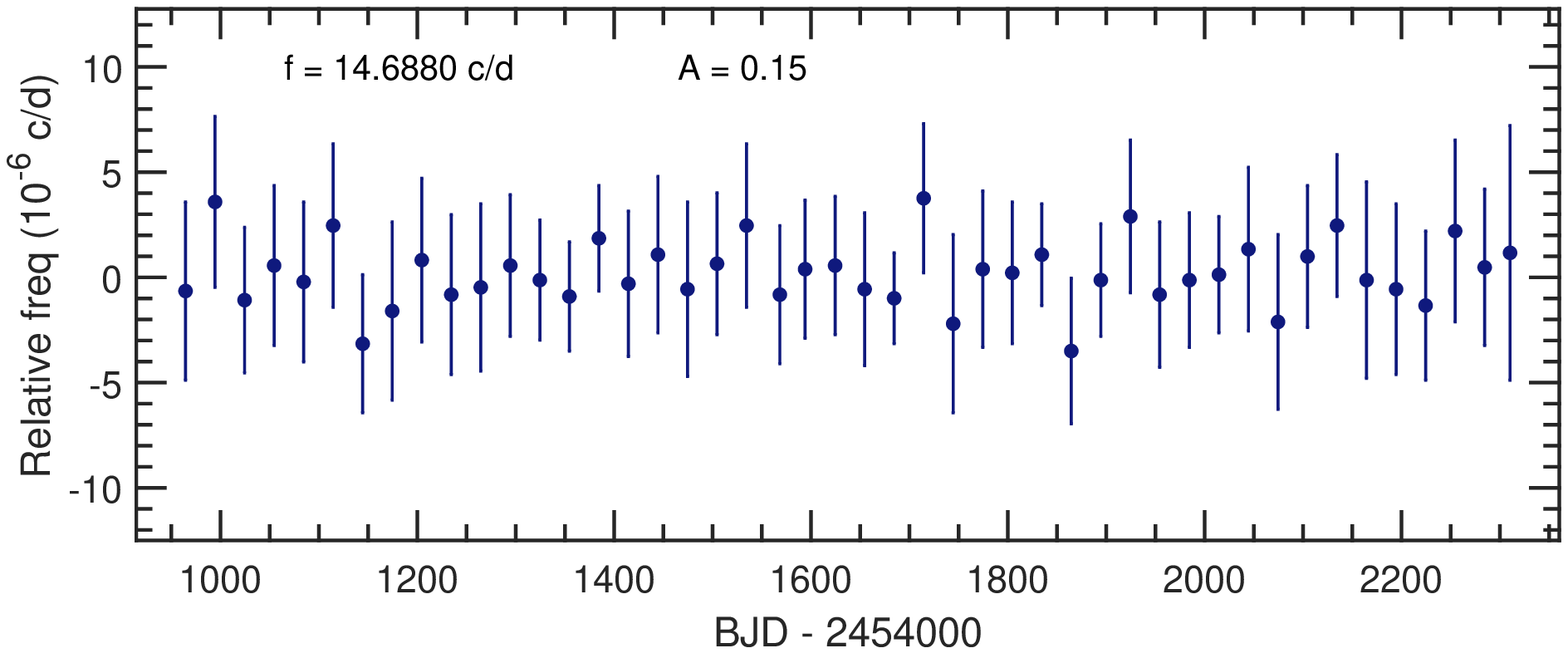}\\
\includegraphics[width=8cm]{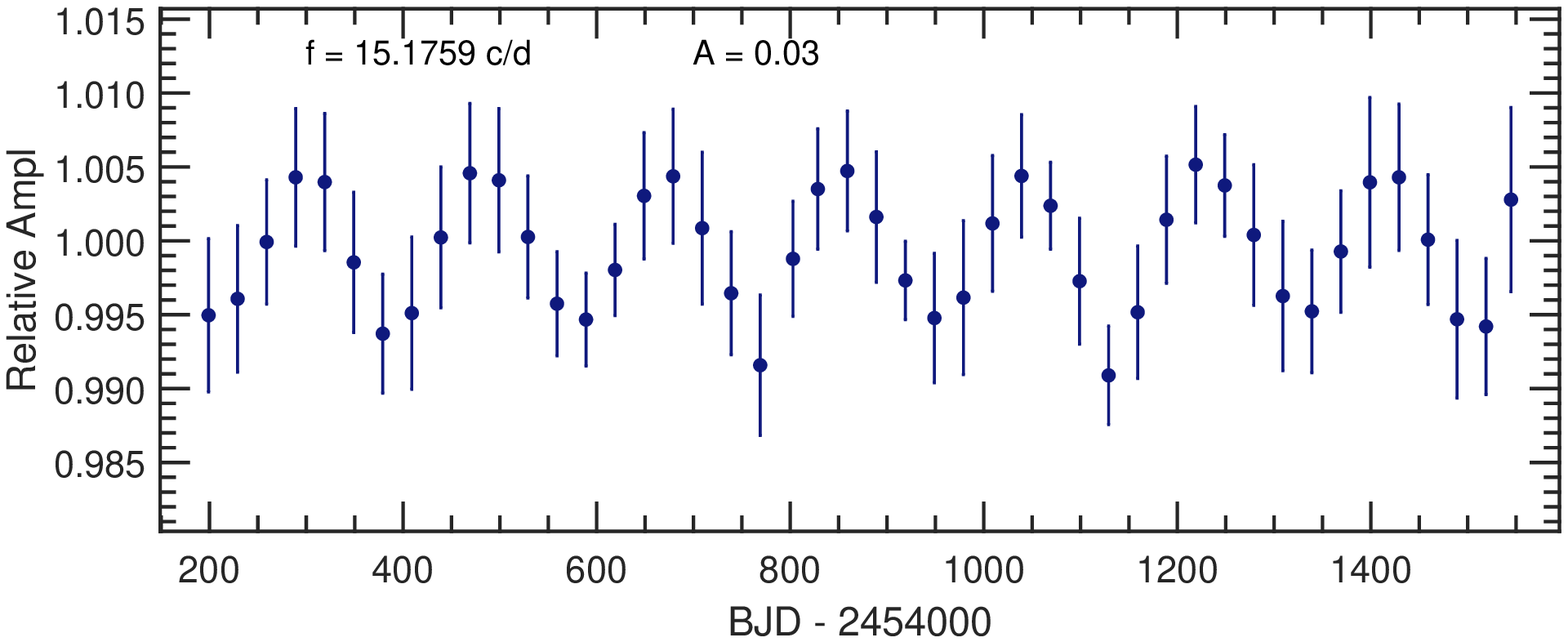}
\includegraphics[width=8cm]{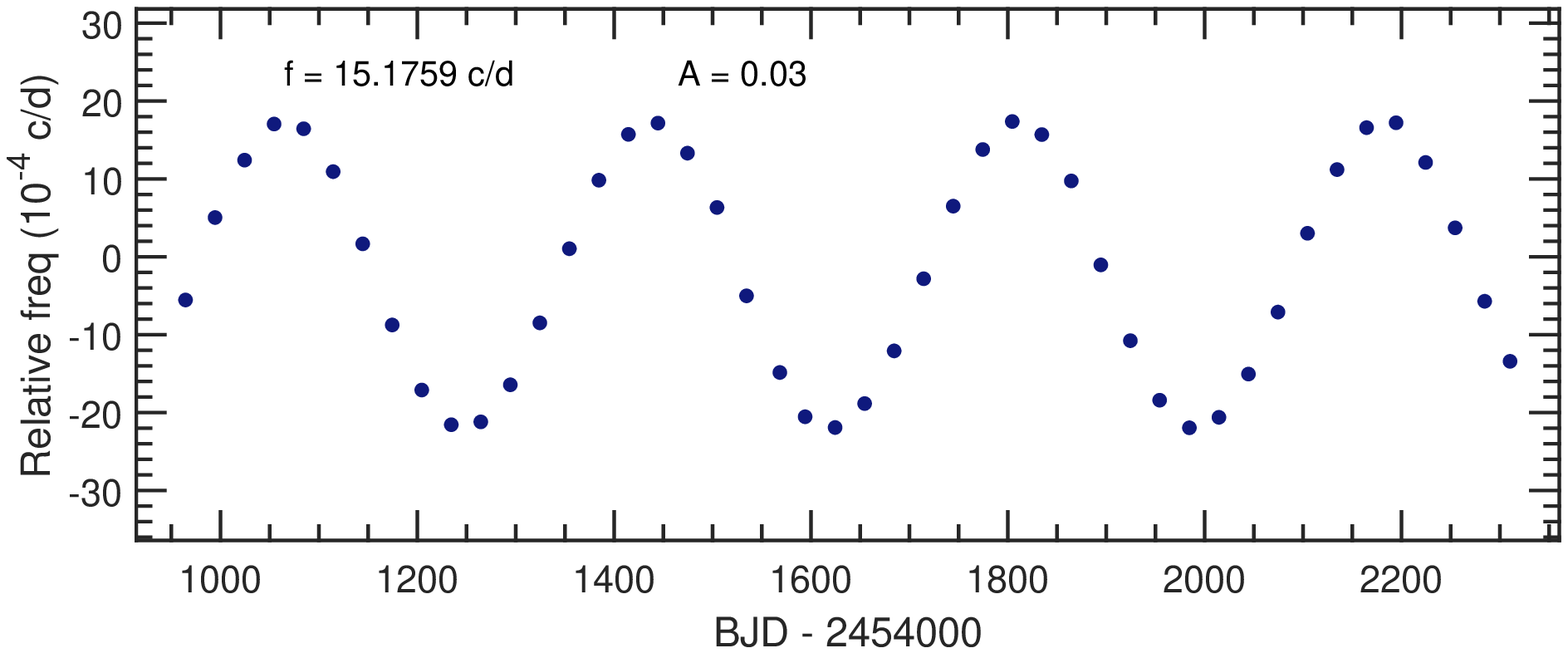}\\
\includegraphics[width=8cm]{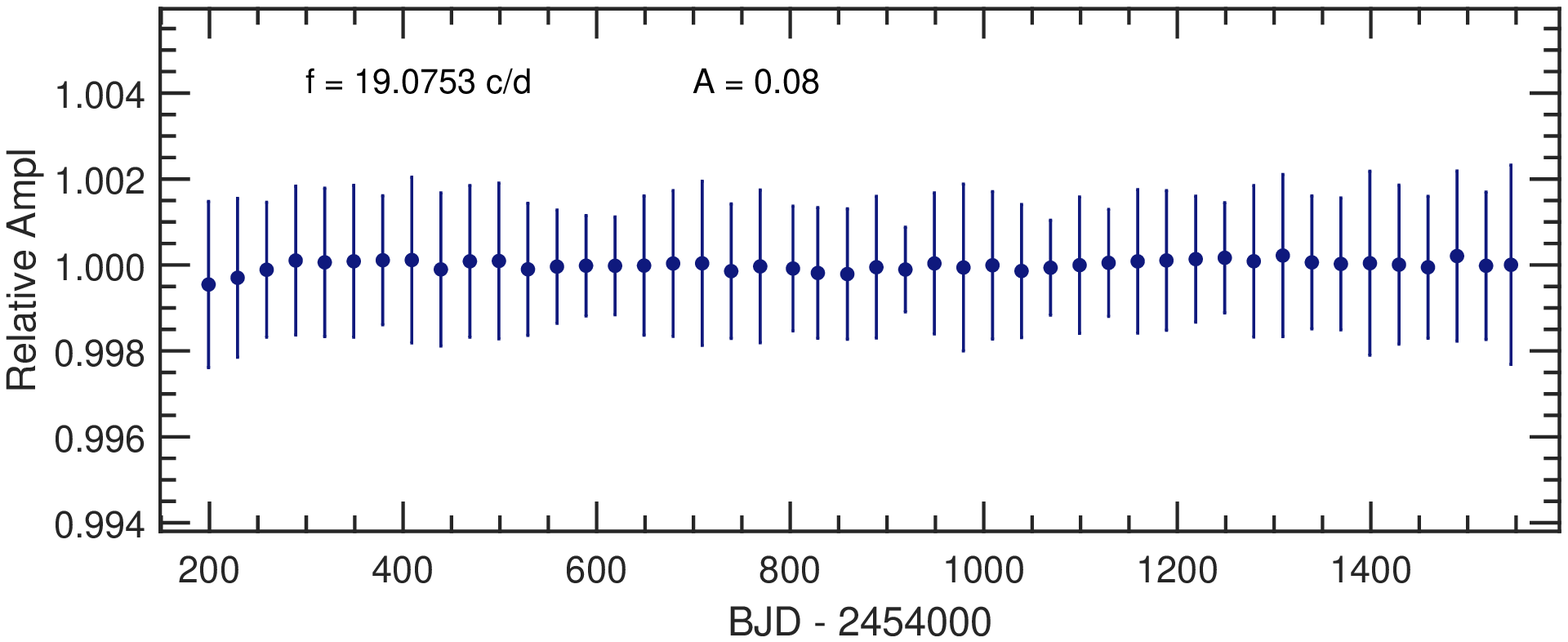}
\includegraphics[width=8cm]{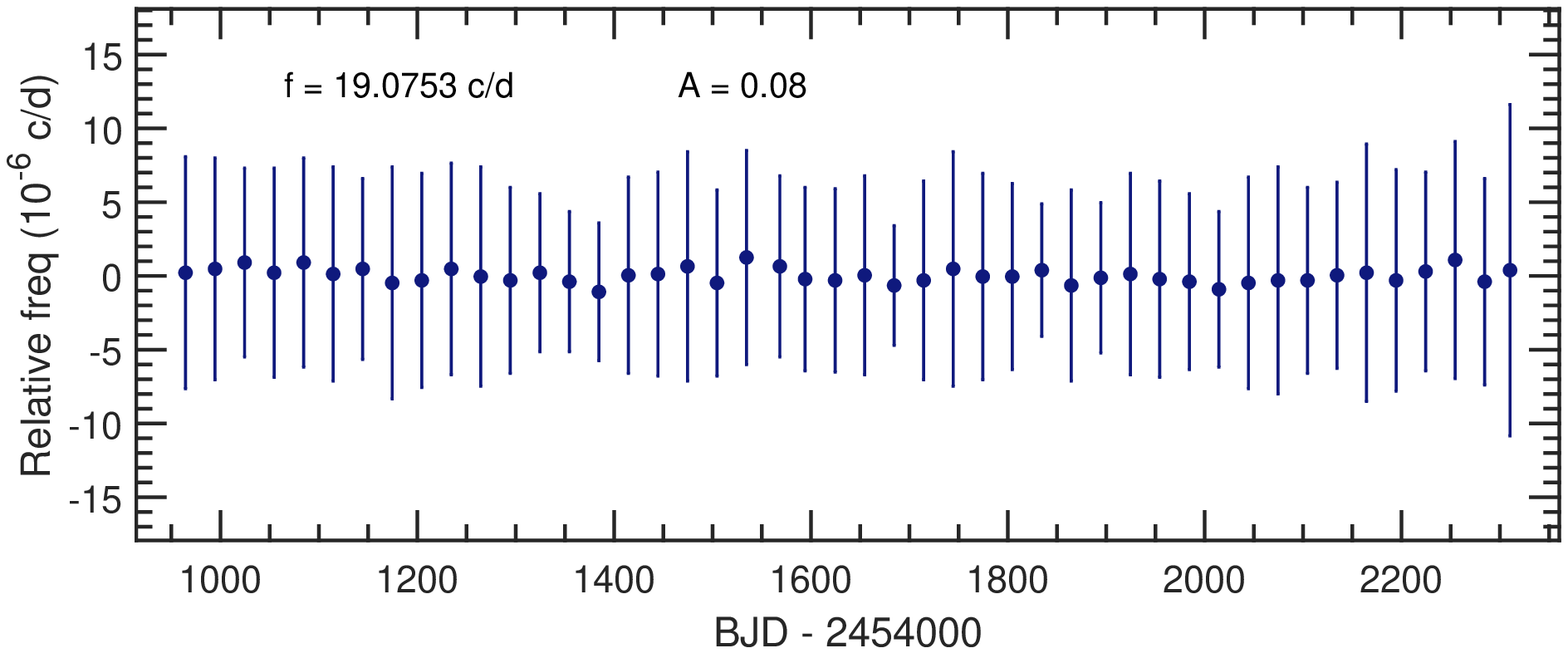}
\caption{Measured amplitude (left panels) and frequency (right panels) variations for the three injected frequencies $f_0$, $f_1$ and $f_2$.
\label{simulation}}
\end{figure*}

Figure\,\ref{simulation} shows the observed modulation patterns for the three injected frequencies. We note that the extracted frequency $f_2 = 2f_{Ny} - f_0 + f_1$ is a mirror reflection relative to the Nyquist frequency, which clearly shows amplitude and frequency modulations with a periodic pattern of about 370~days, i.e., a time scale similar to {\sl Kepler}'s orbital period around the Sun. The amplitude and frequency of $f_2$ varies by about 1\% and 10$^{-3}$~c/d, respectively. In contrast, the frequencies $f_0$ and $f_1$ show no variation in amplitude and frequency. We note that the amplitude and frequency scatter for $f_0$ and $f_1$ are both of the order of 100~ppm and 10$^{-6}$~c/d, respectively, which is smaller than measurement errors.

The frequencies $f_0$, $f_1$, and $f_2$ were all injected with stable amplitudes and frequencies, but they show different modulating characters after extraction. This finding suggests that super-Nyquist frequencies can be identified by amplitude and frequency modulations with a variation of about 1-yr~period. This pseudo modulation is induced by the barycentric time correction, which introduces an unevenly spaced time sampling, with tiny time shifts, in {\sl Kepler}'s data. The Nyquist frequency is the upper limit for equally-spaced data in Fourier transform, beyond which the frequency patterns will repeat again. With barycentric time corrections inducing a variation in the time sampling, the Nyquist frequency changes periodically. Thus the reflection of super-Nyquist signals in the lower frequency range also change following the variation of $f_{Ny}$. We note that amplitude modulations can also be triggered by frequency variations \citep[][see Section~2 for details of mathematical expression]{zong2018}. This kind of variations could therefore easily be used to identify super-Nyquist frequencies in a periodogram. We finally note that these modulations could also generate symmetrical multiplets \citep{2013MNRAS.430.2986M} or complex profiles \citep{2012MNRAS.424.2686B} in Fourier space.

\acknowledgments
We acknowledge support from the National Natural Science Foundation of China through grants 11903005, the Fundamental Research Funds for the Central Universities, and the Agence Nationale de la Recherche (ANR, France) under grant ANR-17-CE31-0018.

\bibliographystyle{aasjournal}

\end{document}